%% file: manuscript.tex
\documentclass{jfm}
\usepackage{graphicx}
\usepackage{epstopdf, epsfig, subfigure}
\usepackage{empheq}
\usepackage{url}

\newcommand{\ie}{i.e.\ }

\newcommand{\Ubulk}{\bar{U}}

\newcommand{\us}{u_s}
\newcommand{\ur}{u_r}
\newcommand{\ut}{u_\theta}

\newcommand{\Ur}{U_r}
\newcommand{\Ut}{U_\theta}

\shorttitle{The vanishing of strong turbulent fronts in bent pipes}
\shortauthor{E.~Rinaldi, J.~Canton and P.~Schlatter}
\title{The vanishing of strong turbulent fronts in bent pipes}

\author{%
	Enrico Rinaldi,
	Jacopo Canton\corresp{\email{jcanton@mech.kth.se}}
	\and Philipp Schlatter
}

\affiliation{%
	Linn\'e FLOW Centre and Swedish e-Science Research Centre (SeRC),
	KTH Mechanics,
	Royal Institute of Technology,
	SE-100 44, Stockholm, Sweden
}

\begin{document}

\maketitle

%-------------------------------------------------------------------------------
\begin{abstract}
\input{abstract}
\end{abstract}


%-------------------------------------------------------------------------------
\section{Introduction}
\label{sec:intro}
\input{introduction}

%-------------------------------------------------------------------------------
\section{Space-time dynamics of localised turbulence}
\label{sec:slugs}
\input{spacetimeq}

%-------------------------------------------------------------------------------
\section{Turbulent kinetic energy budget}
\label{sec:tke_bud}
\input{tkebudget}

%-------------------------------------------------------------------------------
\section{Vortical structures at the front}
\label{sec:vort}
\input{vortical}

%-------------------------------------------------------------------------------
\section{Summary and conclusions}
\label{sec:conclusions}
\input{conclusions}
%

%-------------------------------------------------------------------------------
%\appendix
%\rev{
%\section{Direct-adjoint optimization algorithm}
%\label{sec:optimals}
%\input{appendix}
%}

%===============================================================================
\bibliographystyle{jfm}
\bibliography{bibliography}

\end{document}

%% file: abstract.tex
Isolated patches of turbulence in transitional straight pipes are sustained by a strong
instability at their upstream front, where the production of turbulent kinetic energy (TKE) is up to five times higher than in the core.
%Direct numerical simulations presented in this paper show that this is not the case if the pipe is bent.
Direct numerical simulations presented in this paper show no evidence of such strong fronts if the pipe is bent.
We examine the temporal and spatial evolution of puffs and slugs in a toroidal pipe with pipe-to-torus diameter ratio $\delta=D/d=0.01$ at several subcritical Reynolds numbers.
Results show that the upstream overshoot of TKE production is at most one-and-a-half times the value in the core and that the average cross-flow fluctuations at the front are up to three times lower if compared to a straight pipe, while attaining similar values in the core.
Localised turbulence can be sustained at smaller energies through a redistribution of turbulent fluctuations and vortical structures by the in-plane Dean motion of the mean flow.
This asymmetry determines a strong localisation of TKE production near the outer bend, where linear and nonlinear mechanisms optimally amplify perturbations.
We further observe a substantial reduction of the range of Reynolds numbers for long-lived intermittent turbulence, in agreement with experimental data from the literature.
Moreover, no occurrence of nucleation of spots through splitting could be detected in the range of parameters considered.
Based on the present results, we argue that this mechanism gradually becomes marginal as the curvature of the pipe increases and the transition scenario approaches a dynamical switch from subcritical to supercritical.

%% file: introduction.tex
Over a century ago Osborne Reynolds observed that transition to turbulence in pipe flow is initiated by the appearance of localised travelling patches of chaotic motion~\citep{Reynolds:1883cq}.
He identified as the key parameter for the onset of turbulence the non-dimensional flow rate $\Rey = \Ubulk D / \nu$, later named the Reynolds number, where $\Ubulk$ is the mean flow speed, $D$ is the pipe diameter, and $\nu$ is the kinematic viscosity of the fluid. 
Successive experiments detailed this scenario and drew the distinction between turbulent spots that do not grow in size, \textit{puffs}, and spots that expand in the surrounding laminar flow, \textit{slugs} (see, \eg \citealt{Lindgren:1969dx,Wygnanski:1973fp,Wygnanski:1975fa}; and the review by~\citealt{Mullin2011}).

Pipe flow presents the peculiarity of having a linearly stable laminar velocity profile~\citep{Meseguer:2003kq}, \ie all small perturbations decay and no critical Reynolds number can be defined using linear theory.
However, experiments and simulations show that subcritical transition to turbulence can occur for $Re\gtrsim1700$ if perturbations are sufficiently large.
Only recently a statistical description of the intermittent flow in pipes has provided an accurate estimate of a critical Reynolds number, $\Rey\approx2040$~\citep{Avila2011}.
Below this threshold, the probability of a puff decaying outweighs the probability of  a new puff being generated through a splitting mechanism.
On the other hand, if the Reynolds number is higher than the threshold, the probability of splitting rapidly increases and puffs proliferate.
Theoretical models have been proposed and quantitatively capture this subcritical transition scenario, which falls into the directed percolation universality class~\citep{Barkley:2011kn,Barkley2015,Shih2015}.
For $Re\gtrsim2300$ puffs turn into slugs, which rapidly fill the pipe thereby marking the onset of sustained space-filling turbulence.
The reader is referred to the review by~\cite{Barkley2016a} for further details.

The spatial structure of puffs and slugs has been the subject of several studies initiated by~\citet{Wygnanski:1973fp,Wygnanski:1975fa} and followed by both experimental and numerical investigations (see \eg \citealt{Darbyshire1995,Nishi2008}; and~\citealt{Duguet:2010in,Song2017}, respectively).
Results show a high concentration of turbulent fluctuations at the \textit{strong} upstream front and a gradual decrease at the \textit{weak} downstream front.
In fact, an instability is responsible for extracting energy from the upstream laminar flow and transferring it to the fluctuating field, while the downstream front corresponds to decaying turbulence.
At sufficiently high Reynolds numbers, the weak downstream front turns into a strong front that feeds on the downstream laminar flow~\citep{Song2017}.
The core of the slugs was shown to have the same characteristics of a fully turbulent flow~\citep[see the previously referenced studies and also][]{Cerbus2018}.

\begin{figure}
\centering
\includegraphics[width=0.7\textwidth]{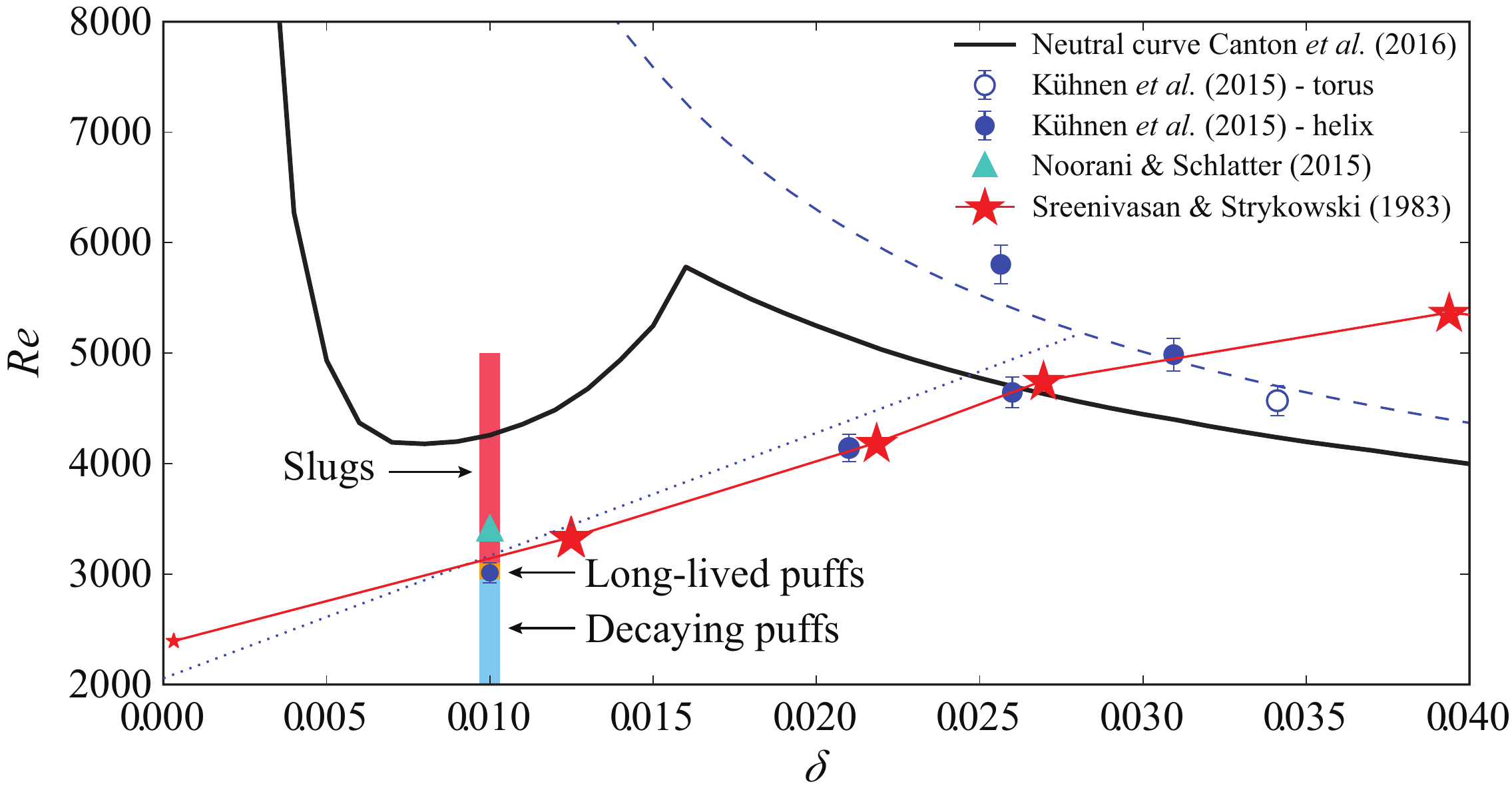}
\caption{Stability map of the flow in a bent pipe. The continuous black line is the neutral curve calculated using linear theory~\citep{Canton:2016in}. Red stars indicate the critical limit for transition as measured by~\citet{Sreenivasan1983}. The triangle is a DNS of a fully turbulent flow by~\citet{Noorani2015}. Circles are experiments by~\citet{Kuhnen2015} indicating $50\%$ intermittency (for $\delta\lesssim0.028$) or the appearance of a periodic travelling wave. The blue lines are the interpolants by~\citet{Kuhnen2015} showing the onset of the subcritical transition (dotted) and of a supercritical bifurcation cascade (dashed).
The coloured boxes indicate the flow regimes characterised in this paper: slugs ($Re>3100$), long-lived puffs ($2950<\Rey<3100$), and decaying puffs ($\Rey<2950$).}
\label{fig:neutral}
\end{figure}

Transition to turbulence in bent pipes has received less attention compared to straight pipes. %, despite the relevance of this geometry.
For this flow case an additional governing parameter comes into play, namely the non-dimensional curvature $\delta=D/d$, with $d$ indicating the coiling diameter.
\cite{White1929} was the first to observe that the flow in a bent pipe can be maintained in a laminar state for higher Reynolds numbers than in a straight pipe.
More recent experiments by \cite{Sreenivasan1983} reported relaminarisation of a turbulent flow entering a coiled pipe and provided Reynolds number ranges for subcritical transition to turbulence for low curvatures.
However, this is not the only transition mechanism in bent pipes.
\cite{Webster:1993ev,Webster1997,Kuhnen2014} have reported supercritical transition from a steady to a periodic flow presenting a travelling wave for medium-high curvatures.
Supercritical transition indicates a qualitative difference between straight and bent pipes, which are characterised by a modal instability for all curvatures different from zero \citep{Canton:2016in}.
A unifying picture for transition in bent pipes was provided by \cite{Kuhnen2015} who reported that subcritical transition dominates for $\delta\lesssim0.028$, while above this value transition to turbulence occurs through a supercritical bifurcation cascade.
Figure~\ref{fig:neutral} reports the neutral curve for the flow in a torus~\citep{Canton:2016in} in the $\delta-\Rey$ plane, and experimental data indicating the onset of turbulence taken from the literature.
Subcritical transition in bent pipes was described as being ``very similar as in straight pipes, where laminar and turbulent flows can coexist'' \citep{Kuhnen2015}.
However, to the best of our knowledge, an in-depth analysis of this regime is still missing.
This is the focus of the present paper.

\begin{figure}
\centering
\includegraphics[width=0.72\textwidth]{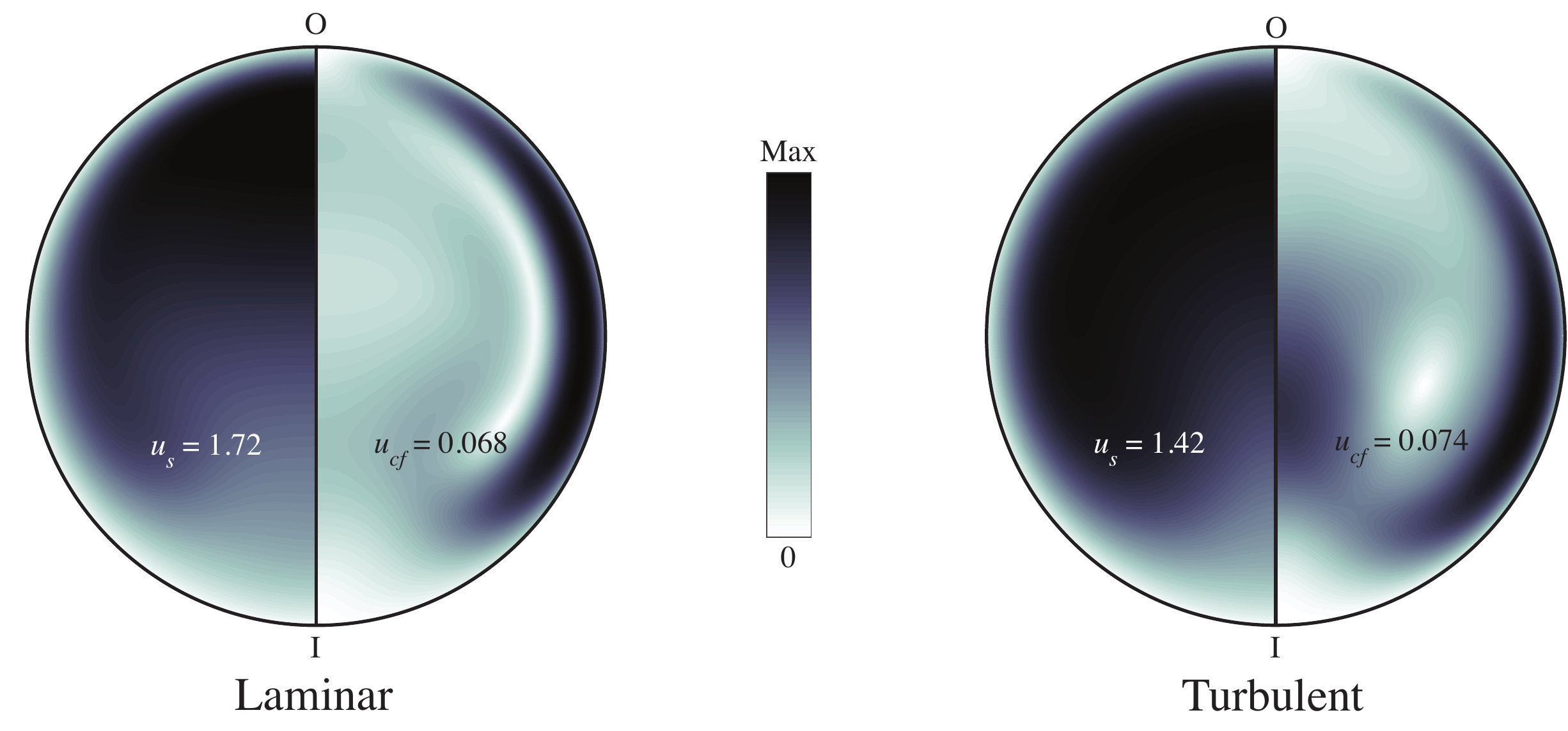}
\caption{Cross-sectional velocity profiles of a laminar (left) and time-averaged turbulent (right) flow at $\Rey=3000$ in a bent pipe with curvature $\delta=0.01$. The inner and outer portions of the bend are indicated by I and O, respectively. The streamwise velocity $u_s$ is displayed on the left half section, while the cross-flow velocity $u_{cf}$ is on the right half section. The maximum values attained on the section are written in the figure.}
\label{fig:base_flow}
\end{figure}

We perform direct numerical simulations (DNS) of the incompressible Navier--Stokes equations
\begin{align*}
&\frac{\p \boldsymbol{u}}{\p t} + \left( \boldsymbol{u} \boldsymbol{\cdot} \boldsymbol{\nabla} \right) \boldsymbol{u} + \boldsymbol{\nabla} p - \frac{1}{\Rey} \nabla^2 \boldsymbol{u} = \boldsymbol{f},\\
&\boldsymbol{\nabla} \boldsymbol{\cdot} \boldsymbol{u} = 0,
\end{align*}
where $\boldsymbol{u}$ is the velocity vector, $p$ is the pressure, $\Rey=UD/\nu$ the Reynolds number (based on bulk velocity and pipe diameter), and $\boldsymbol{f}$ is a body force used to drive the flow in the toroidal geometry~\citep[a detailed discussion of $\boldsymbol{f}$ is given in][]{Canton:2016in,Canton2017}.
Equations are discretized in space and integrated in time using the spectral element code \textsc{nek5000} \citep{nek5000}, which was previously validated on turbulent straight and bent pipes \citep{ElKhoury:2013kj,Noorani2013}.
The flow is kept at unitary bulk velocity as in \citet{Noorani2013,Canton:2016in}.
Exemplary laminar and turbulent velocity profiles in a bent pipe with $\delta=0.01$ and $Re=3000$ are displayed in figure~\ref{fig:base_flow}.
A detailed discussion of how the curvature affects the flow in the laminar regime is given by~\citet{Canton2017}.

In order to ensure that the nature of transition investigated here is
subcritical, we choose $\delta=0.01$.
This curvature is sufficiently lower than the threshold for the onset of a supercritical bifurcation cascade \citep[$\delta\approx0.028$ according to][]{Kuhnen2015}, while introducing a significant deviation of the laminar
flow from the one of a straight pipe \citep{Canton2017}.
At $\delta=0.01$, the laminar flow is linearly unstable for $\Rey>4257$
\citep{Canton:2016in}.
On the other hand, \citet{Kuhnen2015} measured $50\%$ intermittency for $\Rey\approx3000$ and numerical simulations show that the flow is in a sustained fully turbulent state above $\Rey=3400$~\citep[see][]{Noorani2013,Noorani2015}.
Figure~\ref{fig:neutral} highlights the range of Reynolds numbers considered in the present study, $2900\leq\Rey\leq5000$, and gives an overview of our results in terms of turbulent structures.
The large-scale evolution of puffs and slugs is studied in domains of length $L_{s} = 100 D$ and $L_{s} = \pi d/3 \simeq 105 D$ for straight and bent pipes respectively, where the subscript $s$ indicates the streamwise direction.
Our spatial resolution satisfies typical DNS requirements for fully turbulent flows at Reynolds numbers slightly higher than the ones considered here.

The paper continues in \S\ref{sec:slugs} with an analysis of the space-time dynamics of localised turbulent structures.
Section~\ref{sec:tke_bud} presents considerations on the budget of turbulent kinetic energy, and \S\ref{sec:vort} analyses the vortical structures embedded in puffs and slugs.
Section~\ref{sec:conclusions} is dedicated to a discussion of the results.

% turbulent 1.42 and 0.074
% laminar 1.72 and 0.068

%% file: spacetimeq.tex
\begin{figure}
\centering
\includegraphics[width=0.99\textwidth]{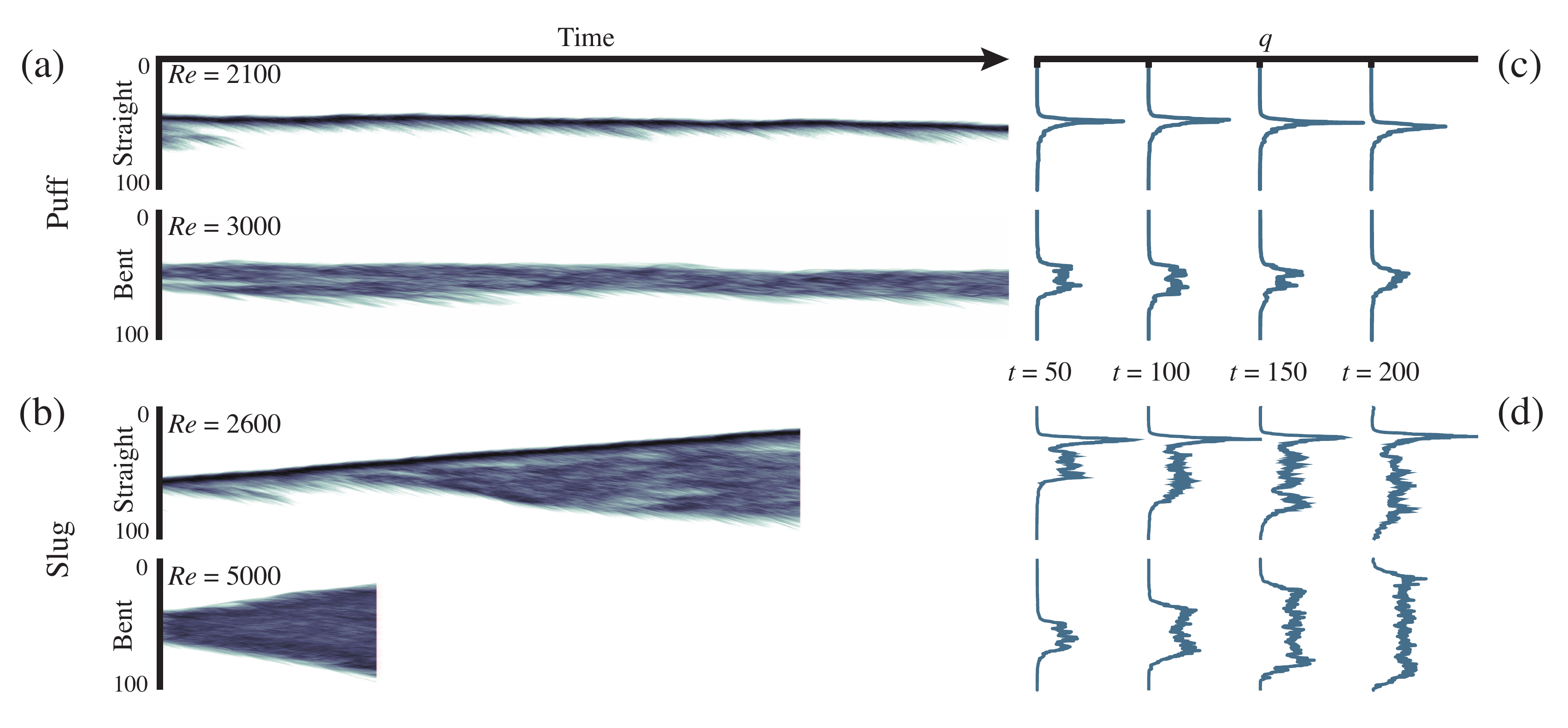}
\caption{Space-time evolution of the cross-flow velocity fluctuations $q$, defined by equation~\eqref{eq:q}, for exemplary puffs (a) and slugs (b) in straight and curved pipes. Colours represent $\log_{10} q$, white corresponds to laminar flow, dark colours to high fluctuations. Panels (c,d) report the spatial distribution of $q$ sampled at four time instants. The (horizontal) scale used to indicate the magnitude of $q$ is the same for the straight and curved pipes.}
\label{fig:spacetime_puffslug}
\end{figure}

We begin by examining how the space--time dynamics of localised turbulence is affected by the curvature.
Figure~\ref{fig:spacetime_puffslug} illustrates the main message of the paper by displaying the evolution of the cross-flow velocity fluctuations, $q$, for puffs (figure~\ref{fig:spacetime_puffslug}(a)) and slugs (figure~\ref{fig:spacetime_puffslug}(b)) in both straight and bent pipes.
We use the scalar quantity $q$ as an indicator of the level of turbulence in accordance with the literature regarding transitional flows in straight pipes \citep[see, \eg][]{Barkley:2011kn} and adapt its definition to the case of bent pipes as
\begin{equation}\label{eq:q}
	q(s,t) = \sqrt{\int_0^{2\pi} \int_0^R \left( \left( \ur - \Ur \right)^2 +  \left( \ut - \Ut \right)^2 \right)r \, \mathrm{dr} \, \mathrm{d\theta} \,}.
\end{equation}
Here, $s$, $r$ and $\theta$ indicate the streamwise, radial and azimuthal directions in toroidal co-ordinates. 
The instantaneous velocity components are $\us = \us (s,r,\theta,t)$, $\ur = \ur (s,r,\theta,t)$ and $\ut = \ut (s,r,\theta,t)$; capital letters denote the laminar flow at a given curvature and Reynolds number.
$\Ur = \Ur (r,\theta; \delta, Re)$ and $\Ut = \Ut (r,\theta; \delta, Re)$ are zero in a straight pipe but not in a curved one, where the
Dean vortices \citep{Dean1927} constitute a secondary motion that can exhibit
intensities comparable to that of the streamwise flow \citep{Canton2017}.
We plot $q = q(s - u_f \, t, \, t)$, calculated in a frame of reference that moves with a constant streamwise velocity $u_f$, and use the same range of colour levels for straight and bent pipes to allow for a direct visual comparison.
The reason for the different Reynolds numbers is that similar flow structures with similar characteristics, \ie isolated puffs moving at approximately constant speed (figure~\ref{fig:spacetime_puffslug}(a)) and slugs expanding at a similar rate (figure~\ref{fig:spacetime_puffslug}(b)), are found at different $\Rey$ in the two pipes.

A clear and distinctive feature differentiates puffs and slugs between the two pipes: the absence of a strong upstream front if the pipe is bent.
Space--time diagrams show the well-known concentration of turbulent fluctuations, indicated by the dark tone of the colour, at the upstream front in straight pipes~\citep{Barkley:2011kn,Barkley2016a,Song2017}.
Conversely, the flow in a bent pipe shows no evidence of this strong front and is characterised by a somewhat uniform distribution of $q$.
This is arguably the result of a gradual process triggered by the presence of a curvature, which leaves no trace of the strong upstream front for $\delta = 0.01$.
A further peculiarity is that puffs extend over a streamwise length that is approximately twice the one reported in the absence of curvature.
One-dimensional profiles of $q$, sampled at several subsequent times, are reported in figures~\ref{fig:spacetime_puffslug}(c) and (d), and facilitate the comparison between the straight and bent pipe flow cases.
The overshoot of turbulent fluctuations is prominent if the pipe is straight and can reach values significantly higher than the ones downstream of the front, 
while in the curved pipe the upstream front has intensity comparable to that of the core of the structure.

\begin{figure}
\centering
\includegraphics[width=0.5\textwidth]{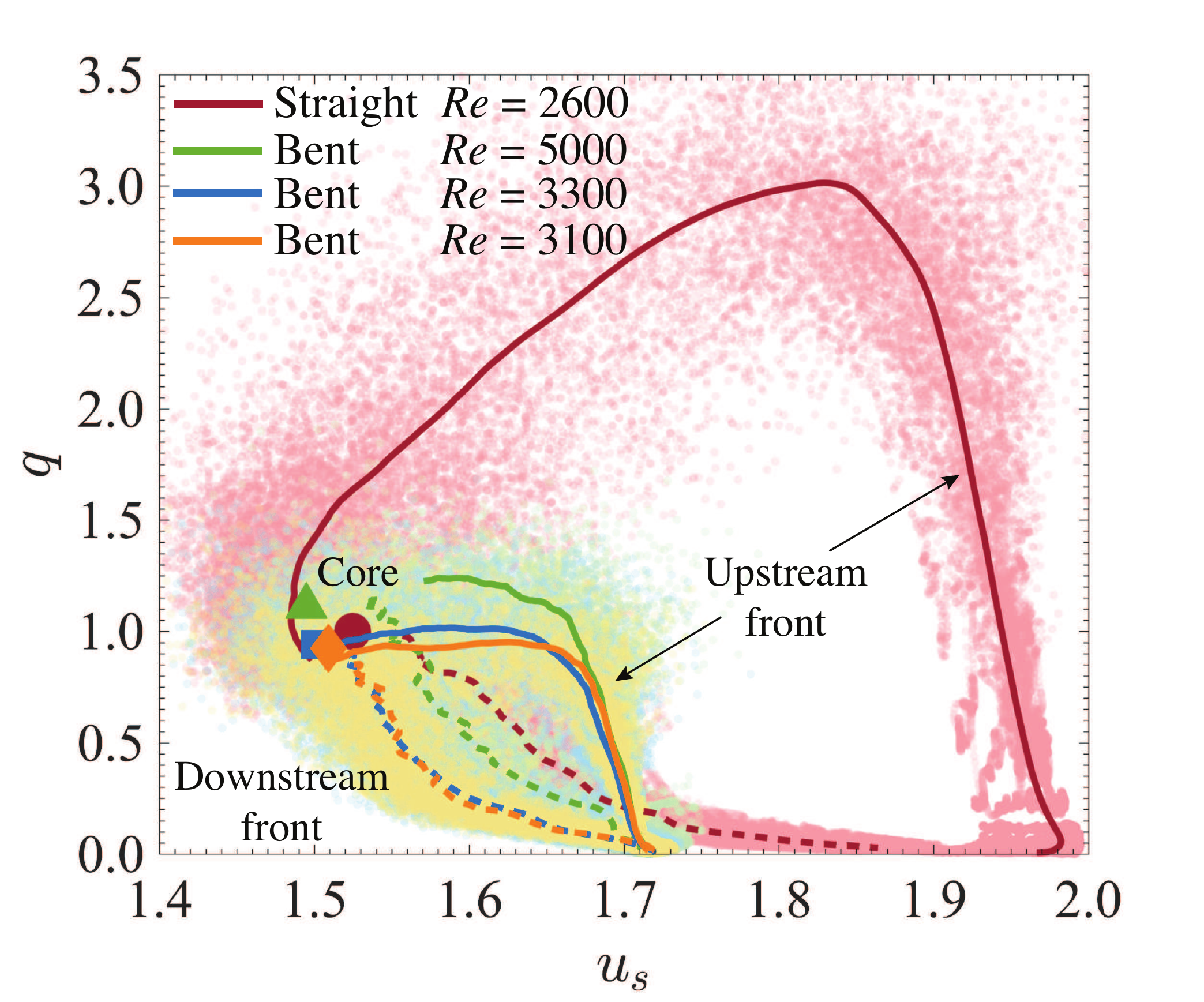}
\caption{Trajectories of turbulent slugs in the $\us-q$ phase space for straight and curved pipes. The clouds indicate a subset of the instantaneous values. Lines indicate the average over time of values at the upstream front (solid) and at the downstream front (dashed). Large filled symbols are space-time averages in the core of the slug.
Values of $q$ are normalised by the average turbulent fluctuations in the core of the slug in the straight pipe at $\Rey=2600$.}
\label{fig:phase_space}
\end{figure}

The structure of puffs and slugs is quantitatively characterised by means of a projection of the flow trajectory on the $u_s - q$ phase space~\citep{Song2017}, with $u_s$ now being the maximum streamwise velocity on the cross-section on which $q$ is calculated.
Figure~\ref{fig:phase_space} presents the data for a single realisation of a slug in a straight pipe at $\Rey=2600$ and three slugs in a bent pipe at $\Rey=3100$, 3300 and 5000.
The loops for each flow case are obtained by time averaging $u_s - q$ pairs over a length of 20 diameters across the upstream and the downstream fronts. 
Large filled symbols are space--time averages of the trajectory in the core.
All values of $q$ are normalised by the average turbulent fluctuations in the core of the slug in the straight pipe at $\Rey=2600$.
Moving in the direction of the flow, which corresponds to following loops in anti-clockwise direction starting from $q=0$, the laminar velocity profile flattens, which is indicated by a decrease of $\us$, and the turbulent fluctuations rapidly increase.
The overshoot of $q$ is clearly visible in the case of straight pipe, where maximum values can be over three times larger than in the core.
On the other hand, if the pipe is bent, $q$ settles to approximately the same value attained in the core in a nearly monotonic fashion.

The steep increase of $q$ is a common feature of the upstream fronts in straight and bent pipes, and indicates that a similar mechanism of instability sustains the slug by extracting energy from the surrounding laminar flow.
This is a key similarity that suggests that the analogy between transitional pipe flow and an excitable medium~\citep{Barkley2012}, such as a nerve axon, still holds in its essence if the pipe is bent.
However, our results indicate a significant reduction of super-threshold perturbations at the upstream front as the curvature increases.
This goes hand in hand with the emergence of strong Dean vortices that enhance the mixing and redistribution of turbulent fluctuations (the supplementary video 1 gives a graphical representation of the action of Dean vortices).
As opposed to the upstream front, the downstream front corresponds to decaying turbulence, as in straight pipes.

The features of puffs and slugs discussed above are not affected by the choice of the parameter indicating the level of turbulence.
The definition of $q$ reduces the velocity field to a single scalar value by integrating over the cross-section of the pipe.
This is a reasonable approximation in the case of a straight pipe, where the azimuthal direction is homogeneous. Nevertheless, $q$ can still underestimate the level of turbulence in the presence of strong localisation in the radial direction.
In the case of a bent pipe further localisation can occur in the azimuthal direction, which is not homogeneous due to the curvature.
We have verified that replacing (the average) $q$ with its local maxima results in the same monotonic increase towards the core with no trace of overshoot.

\begin{figure}
\centering
\includegraphics[width=0.99\textwidth]{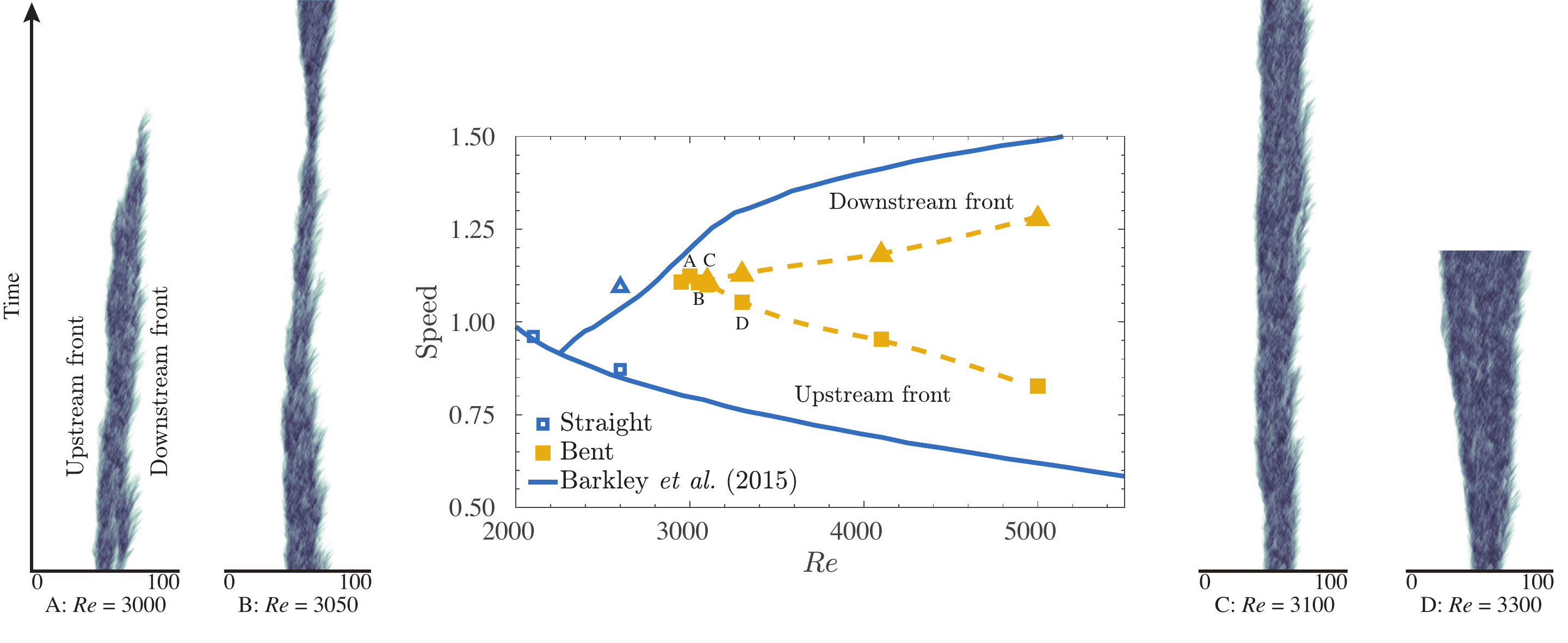}
\caption{Front speed in straight (empty symbols) and bent pipes (filled symbols) as a function of the Reynolds number. Squares and triangles indicate the upstream and downstream fronts, respectively.
At a given $\Rey$, a single entry in speed identifies a puff.
The continuous line is the model from~\citet{Barkley2015} for straight pipe. 
The dashed line guides the eye through the bent pipe data, but does not represent a model for the speed of the fronts.
The lateral panels show the space--time diagrams corresponding to four selected cases. The time scale (vertical axis) is the same for all panels; the puff at $\Rey=3000$ decays after $\tau\approx700 \, D/U$.}
\label{fig:speed}
\end{figure}

Figure~\ref{fig:speed} presents a parametric study of the effect of the Reynolds number over the evolution of localised turbulence in the $\delta=0.01$ bent pipe.
The image includes space-time diagrams of $q$ and an estimate of the velocity of the upstream and downstream fronts.
The aim is to define an approximate lower bound on $\Rey$ for long-lived localised turbulence in the form of puffs, and to pinpoint the onset of expanding turbulence.
For our purpose, we set the definition of long-lived turbulence to a decaying time  $\tau > 1000 \, D/U$, which is shorter than the typical values used in straight pipes. As we will discuss in the following paragraphs, in bent pipes the separation in Reynolds number between the puff and slug regimes is much smaller than in straight pipes \citep{Sreenivasan1983}. This indicates that for similar accuracy in $\Rey$ shorter survival times for long-lived localised turbulent structures are sufficient.

Our results indicate that for $\delta=0.01$ turbulence decays for Reynolds numbers below $2950$.
Isolated puffs seeded into a steady flow relaminarise, as do initially fully turbulent flows computed for higher Reynolds numbers, following a protocol analogous to~\cite{Moxey2010a}.
At $\Rey=3000$ a single realisation of an isolated puff decays in approximately 700 advection time units (see point A in figure~\ref{fig:speed}).
On the other hand, a flow initialised with a fully turbulent field appears to persist as such.
At $\Rey=3050$ two simulations initialised with localised turbulence show no return to a laminar state.
The first separation between the upstream and downstream fronts speed (approximately 1\%) occurs at $\Rey=3100$ and marks the onset of the slug regime.

The limited number of realisations presented in this paper does not allow a precise computation of front speeds and decay times.
Nonetheless, figure~\ref{fig:speed} provides an overview of the subcritical transition scenario in bent pipes and how it differs from straight ones.
We document an overall increase of the Reynolds numbers at which subcritical transition is triggered.
The range of values at which long-lived localised turbulence persists is also significantly narrowed, reducing from $1900<\Rey<2300$ for a straight pipe to $2950<\Rey<3100$, above which puffs turn into slugs.
The same behaviour was observed by~\citet{Sreenivasan1983} with experiments on helical pipes, and our numerical results are in good agreement with their data and with the $\Rey$ value indicated by \cite{Kuhnen2015} for 50\% intermittency.

The Reynolds number is not the only quantity to increase for bent pipes as compared to straight ones: the mean advection velocity of puffs and slugs is also larger.
Puffs are faster than the bulk flow, while in a straight pipe they are slower, and the upstream front of slugs is slower than the mean flow only for $\Rey\gtrsim3600$.
Conversely, the expansion rate of slugs is reduced when compared to a straight pipe, indicating that these structures travel downstream more quickly but perturb a stationary point for a shorter time.

%% file: tkebudget.tex
\begin{figure}
	\centering
	\includegraphics[width=0.99\textwidth]{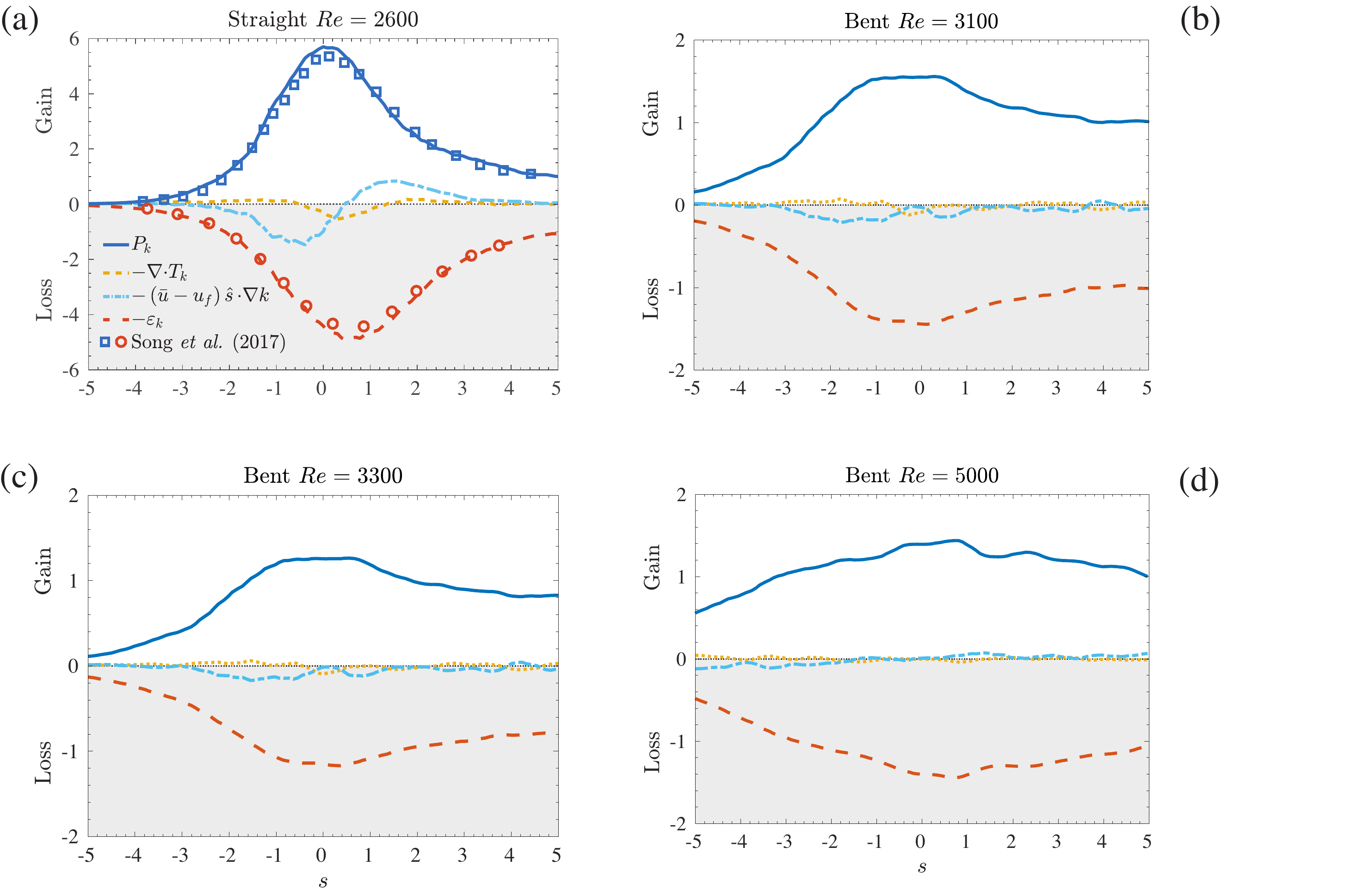}
	\caption{%
		Turbulent kinetic energy budget integrated over the pipe
		cross-section in a straight pipe (a) and in a bent pipe for different
		Reynolds numbers (b)--(d).
		Panel (a) also reports the data by \cite{Song2017}.
		The budget is calculated in a moving frame of reference centered at the
		upstream front of the slug.
		The terms of the budget are normalised by the value that $P_k$ assumes in
		the core of the slug.
		Note the different scaling of the vertical axes.
	}
	\label{fig:tke_budget}
\end{figure}

In order to get further insight into the structure of localised turbulence
in bent pipes, figure~\ref{fig:tke_budget} presents the budget
of turbulent kinetic energy $k= 1/2 \, \overline{u'_i u'_i}$ calculated in the frame of reference co-moving with a slug and centered on the upstream front.
The overline indicate time-averaged quantities and primes the fluctuation field $u'_i = u_i - \overline{u_i}$.
Figure~\ref{fig:tke_budget}(a) presents the budget for a straight pipe at $\Rey=2600$ and a comparison with data by \cite{Song2017}.
Figure~\ref{fig:tke_budget}(b)-(d) shows the present results in bent pipes for $\Rey=3100$, $3300$ and $5000$.
In the relative frame of reference the budget reads
\begin{equation}
	\frac{\partial k}{\partial t} = P_k - \varepsilon_k - \left( \bar{u} - u_f \right) \boldsymbol{\hat{s}} \boldsymbol{\cdot} \boldsymbol{\nabla} k - \boldsymbol{\nabla}\! \boldsymbol{\cdot} \boldsymbol{T}_k = 0, \label{eq:tke}
\end{equation}
%
%\begin{subequations}
	\begin{equation}
		P_k = - \overline{u'_i u'_j} \frac{\partial \overline{u_i}}{\partial x_j},
		\qquad
		\varepsilon_k = \frac{2}{\Rey} \overline{ s_{ij} s_{ij} },
		\qquad
		T_{k,i} = \frac{1}{2} \overline{ u'_i u'_j u'_j } + \overline{ u'_i p' } - \frac{2}{\Rey}\overline{ u'_j s_{ij} },
	\end{equation}
%\end{subequations}
%
where $s_{ij} = 1/2 \left( \partial u'_i / \partial x_j + \partial u'_j / \partial x_i \right)$ indicates the shear rate, $p$ is the pressure and $\boldsymbol{\hat{s}}$ is the unit vector pointing in the streamwise direction. $P_k$, $\varepsilon_k$ and $T_{k,i}$ are the production, dissipation and transport of turbulent kinetic energy, respectively.
The terms in equation~\eqref{eq:tke} are averaged over the pipe cross-section
and normalised by the production rate in the core of each slug.
As for $q$ in the previous section, we have verified that using mean or local values does not affect the results.

The most striking difference between straight and bent pipes is the value of $P_k$ and $\varepsilon_k$ at the upstream front.
While in a straight pipe the average production at the front is approximately five times higher than in the core, in a bent pipe it is only one-and-a-half times the value of the core.
The second relevant difference is that in a straight pipe the peak of production preceeds the peak of dissipation by about one pipe diameter, as can be seen in figure~\ref{fig:tke_budget}(a), while in a bent pipe they are almost at the same location (see panels (b)-(d)).
Since for slugs in a bent pipe $P_k$ and $\varepsilon_k$ are in near equilibrium at each streamwise location, the rates of energy flux and convection provide a minor contribution to the budget.

\begin{figure}
\centering
\includegraphics[width=0.85\textwidth]{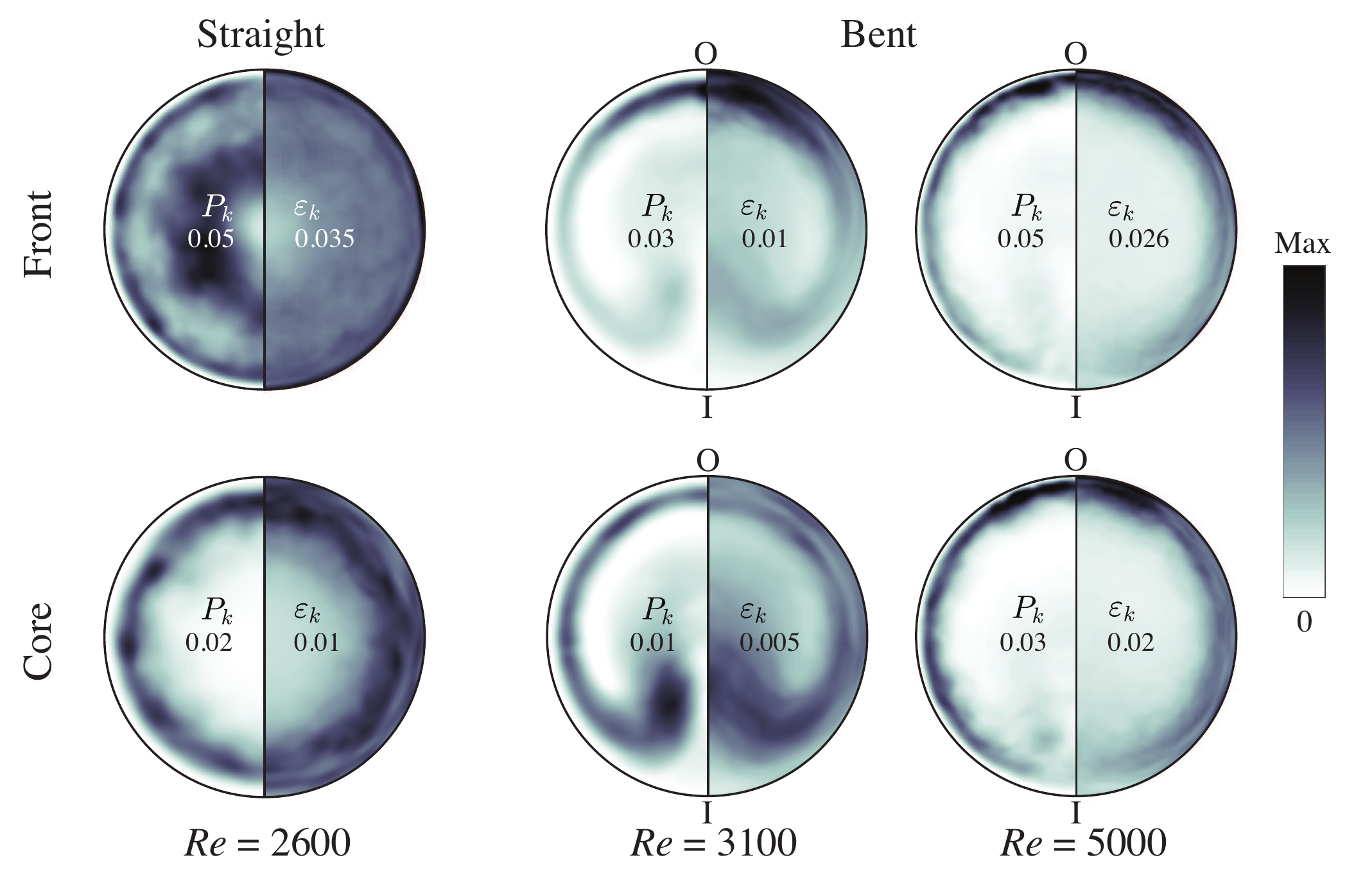}
	\caption{%
		Turbulent kinetic energy production (left halves) and dissipation (right
		halves) at the front and core of slugs in straight and bent pipes.
		All quantities are averaged over time by tracking the slug, and over the
		cross-section taking into account the mirror symmetry of the torus,
		the axial invariance of the pipe has not been used in order not to alter
		the comparison.
		The text labels report the maxima of $P_k$ and
		$\varepsilon_k$ on each section (see also figure~\ref{fig:tke_budget}).
		The outer and inner portions of the bend are indicated by O and I, respectively.
	}
	\label{fig:2d_budget}
\end{figure}

One-dimensional budgets provide only partial information as bent pipes are not invariant to rotations on the section.
As reported by \cite{Sreenivasan1983} turbulent fluctuations in bent pipes appear first in the outer portion of the bend, while they pervade the whole cross-section only for higher Reynolds numbers.
To complement figure~\ref{fig:tke_budget}, figure~\ref{fig:2d_budget} presents the time averaged distributions of production and dissipation over a cross-section of the pipe.

The panels in the top row of figure~\ref{fig:2d_budget} are computed at the upstream front of slugs, which corresponds to $s=0$ in figure~\ref{fig:tke_budget}.
In a straight pipe $P_k$ and $\varepsilon_k$ are mainly concentrated in the near-wall region and in a ring around the centre of the pipe.
Conversely, the budget in a bent pipe shows a high localisation towards the outside of the bend and lower peak values.
This confirms that the vanishing of the strong upstream front is not an artifact of spatial averaging, but a specific feature pertaining to bent pipes.
The cross-sectional localisation and lower local values of $P_k$ and $\varepsilon_k$ also suggest that an additional mechanism must come into play in sustaining turbulent patches.
We argue that the secondary motion created by the Dean vortices is at the root of both these phenomena.

An insight in the spatial distribution of $P_k$ at the front is given by considering perturbations whose integrated squared velocity is optimally amplified in the linear (small amplitude) and nonlinear (finite amplitude) regime, for a given time horizon. 
The structures corresponding to these optimal perturbations identify highly receptive regions of the flow where instabilities are likely to be triggered and lead to transition to turbulence.

In a straight pipe the linear optimal perturbation consists in a pair of large-scale low- and high-speed streaks, and corresponding counter-rotating vortices, which are located in the centre of the pipe. The nonlinear optimal perturbation, instead, is formed by streamwise and azimuthally localised streaks near the wall \citep[see, e.g.][]{Pringle2010}.
The position of linear and nonlinear optimals corresponds to the peaks of turbulent kinetic energy production at the upstream front of puffs, thus suggesting a connection with the instability mechanisms that sustain the structure.

\begin{figure}
	\centering
	\includegraphics[width=0.85\textwidth]{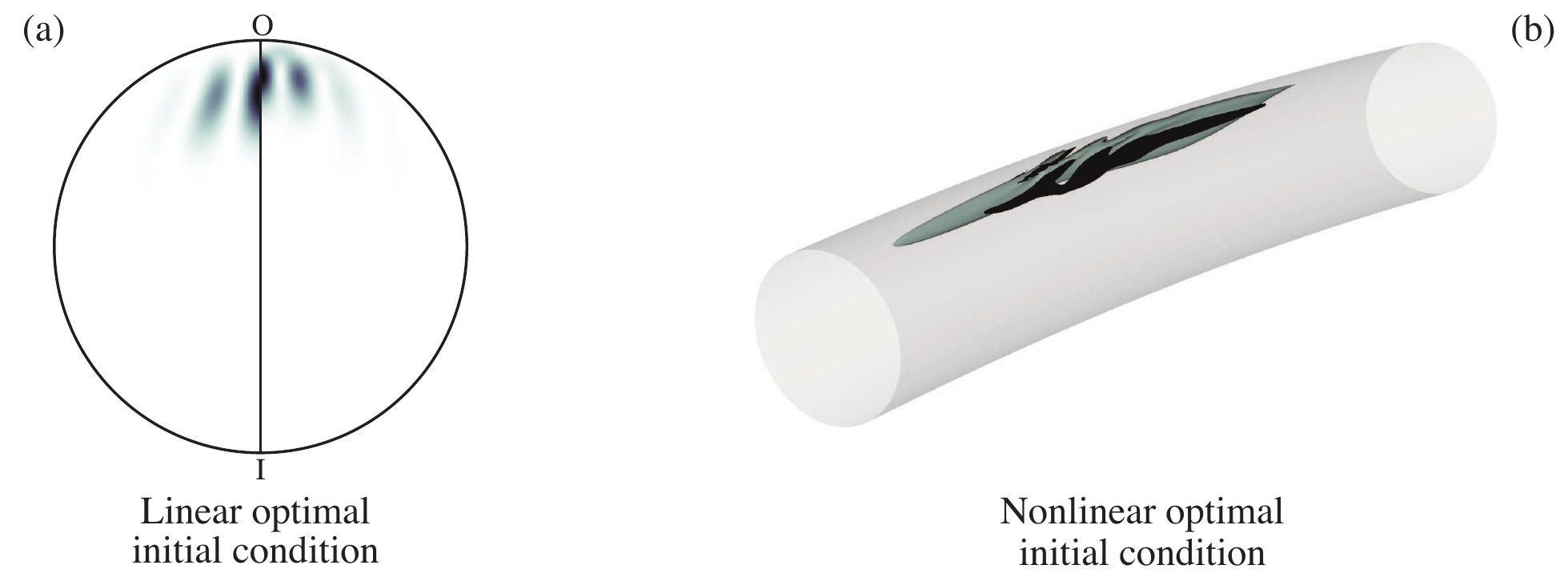}
	\caption{%
	Optimal perturbations in a bent pipe. (a) Linear optimal initial condition: the left and right
	half of the pipe section depict $E_{sw} = \frac{1}{2}u_s^2$ and $E_{cf} = \frac{1}{2} \left(u_r^2 + u_\theta^2 \right)$,
	respectively.
	(b) Nonlinear optimal initial condition: contours of the streamwise velocity ($-3.5\times
	10^{-3} U$ and $2.5\times 10^{-3} U$).
	}
	\label{fig:optimals}
\end{figure}

We have performed an analogous analysis for a bent pipe with $\delta = 0.01$ and present the results in figure~\ref{fig:optimals}.
Linear optimals were calculated using the finite element code \textsc{PaStA}~\citep{Canton2013,Canton:2016in}.
The figure shows the optimal perturbation calculated at $\Rey = 2870$, which is characterized by an energy gain of $106$ by $t = 7.5$ with respect to the value at $t=0$.
The spatial distribution of the initial velocity field shows a strong cross-sectional localisation near the outer portion of the bend.
In this region the streamwise velocity of the laminar base flow as well as its wall-normal derivative are largest, suggesting a large potential for extracting kinetic energy from the steady flow and converting it into the fluctuating field.
The streamwise streaks are quickly amplified through the lift-up induced by the counter-rotating vortices and drift to the sides of the section as a consequence of the secondary motion induced by the Dean vortices, dissipating as they move away from the outer portion of the bend, and eventually disappearing entirely by the time they reach the inner bend.

Nonlinear optimal perturbations were calculated with \textsc{nek5000} using a gradient-based optimisation algorithm that relies on integrating backward in time the adjoint equations to compute the gradient~\citep{Pringle2010}.
A checkpointing and revolve strategy was adopted in order to limit the memory requirements~\citep{Schanen2016}.
The initial condition is updated at each step of the optimization using the gradient rotation algorithm~\citep{2013_FouresEtAl}.
The nonlinear optimal perturbation at $\Rey=3000$ with initial energy constrained to $E_0 = 4\times 10^{-5} U^2$ (integrated over the volume) is shown in the figure.
The length of the domain is $L_s = 2\pi D$.
The optimization was performed over a maximum integration time $T = 60 D/U$, at which the energy gain of the converged optimal is $1074$.
Similarly to the linear optimal, the streaky velocity perturbations are localised near the outer bend. They undergo secondary instabilities leading to the transient appearance of hairpin vortical structures and are also subjected to the drift induced by the Dean vortices.
Despite the significant energy amplification of the reported nonlinear optimal perturbation, its evolution beyond the optimization time $T = 60 D/U$ is marked by a decay and an eventual return of the flow to a laminar state.
We have not searched for a minimal seed of turbulence, which implies a search over both $E_0$ and $T$ parameter spaces and goes beyond the scope of the discussion of optimal perturbations within the present paper.

From the results presented it appears that the high receptivity of the region near the outer portion of the bend, where the recirculating mean flow impinges on the wall of the pipe, plays a primary role in feeding the upstream front of the puff and keeping it alive, as suggested by the location of the peak of $P_k$.
Turbulent fluctuations generated here are then transported along the walls of the pipe and lifted up towards the inner section.

The bottom row of figure~\ref{fig:2d_budget} presents the distribution of production and dissipation at the core of the slugs.
In a straight pipe the ring of high $P_k$, which was observed at the front, disappears.
Production and dissipation are localised in the near-wall region and are the same as in a fully turbulent pipe flow.
For high Reynolds numbers, when slugs expand rapidly, the spatial distribution of $P_k$ and $\varepsilon_k$ in bent pipes is qualitatively unchanged between the front and the core, as can be seen in the rightmost column of figure~\ref{fig:2d_budget}.
At intermediate Reynolds numbers, instead, the cores are qualitatively different.
The peaks of $P_k$ and $\varepsilon_k$ are localised in two lobes in the proximity of the inner bend of the pipe, near the symmetry axis, as can be seen in the central column of figure~\ref{fig:2d_budget} for $\Rey=3100$.
The same behaviour is observed for $\Rey = 3300$ (not shown in the figure).
These lobes are likely connected with a low frequency, highly energetic mode responsible for the so-called sublaminar drag in the range $2900\lesssim\Rey\lesssim4000$ \citep{Noorani2015}.

%% file: vortical.tex
\begin{figure}
	\centering
	\includegraphics[width=\textwidth]{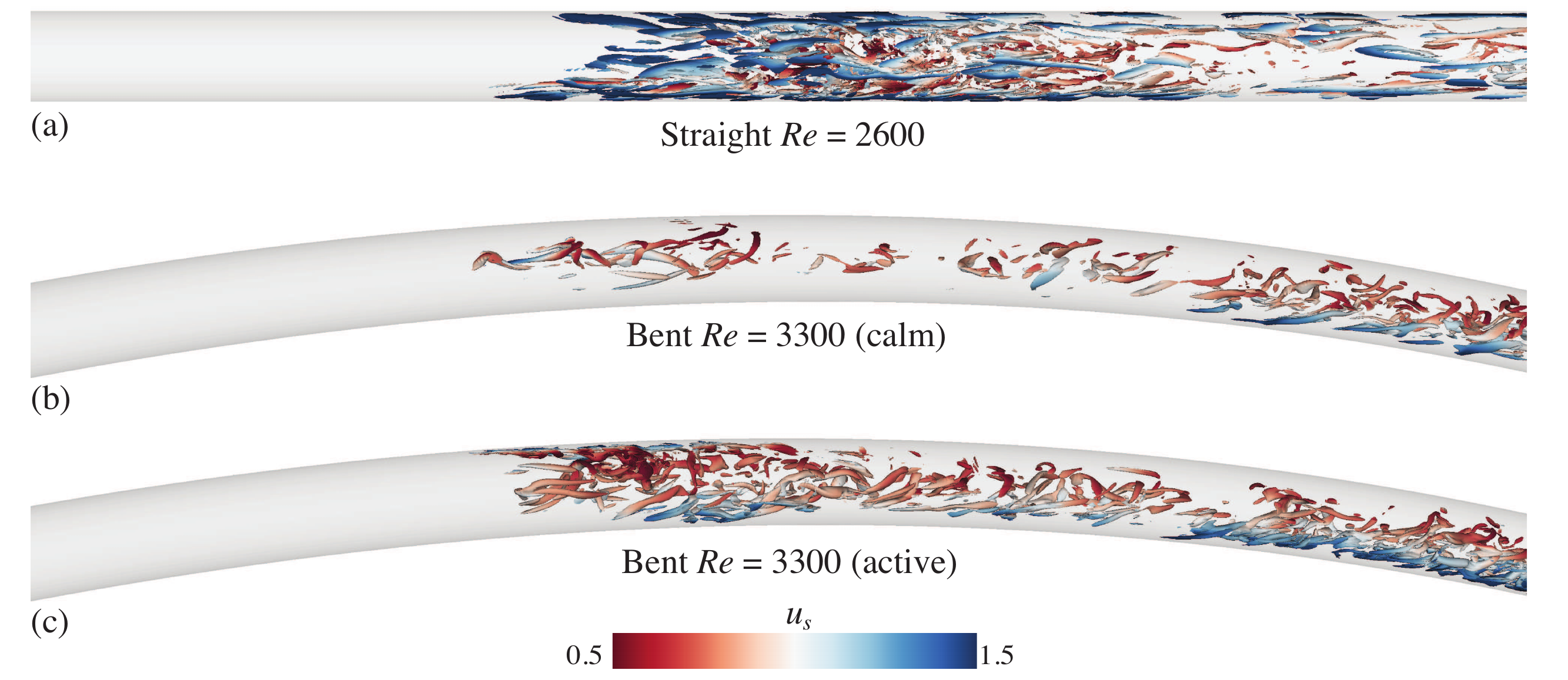}
	\caption{%
		Snapshots of the flow at the upstream front of a slug.
		The figure depicts iso-contours of negative $\lambda_2$ coloured by
		streamwise velocity magnitude.
		The slug in the bent pipe is represented in both its ``calm'' and
		``active'' phases.
		See also the supplemental videos 1-4.
	}
	\label{fig:lambda2}
\end{figure}

We conclude our analysis by discussing the dynamics of vortical structures at the upstream front of slugs.
Figure~\ref{fig:lambda2} reports instantaneous fields of vortices visualised using the $\lambda_2$ criterion \citep{Jeong:1995ep} for both straight and bent pipes.
This method identifies the vortex core using the second eigenvalue of the symmetric tensor $S^2 + \Omega^2$, with $S$ and $\Omega$ indicating the symmetric and antisymmetric parts of the velocity gradient tensor.

The flow in a straight pipe shows a high concentration of vortices at the upstream front, distributed in the typical conical shape going from the near-wall region towards the centre (figure~\ref{fig:lambda2}(a)).
These vortices form a relatively sharp boundary with the laminar flow upstream of the front and are generated at an approximately constant rate, as can be seen in the supplementary video 2.
Their concentration in the centre of the pipe reduces downstream of the front and is limited to the near wall region, consistently with the distribution of $P_k$ discussed in \S\ref{sec:tke_bud}.

\begin{figure}
	\centering
	\includegraphics[width=0.65\textwidth]{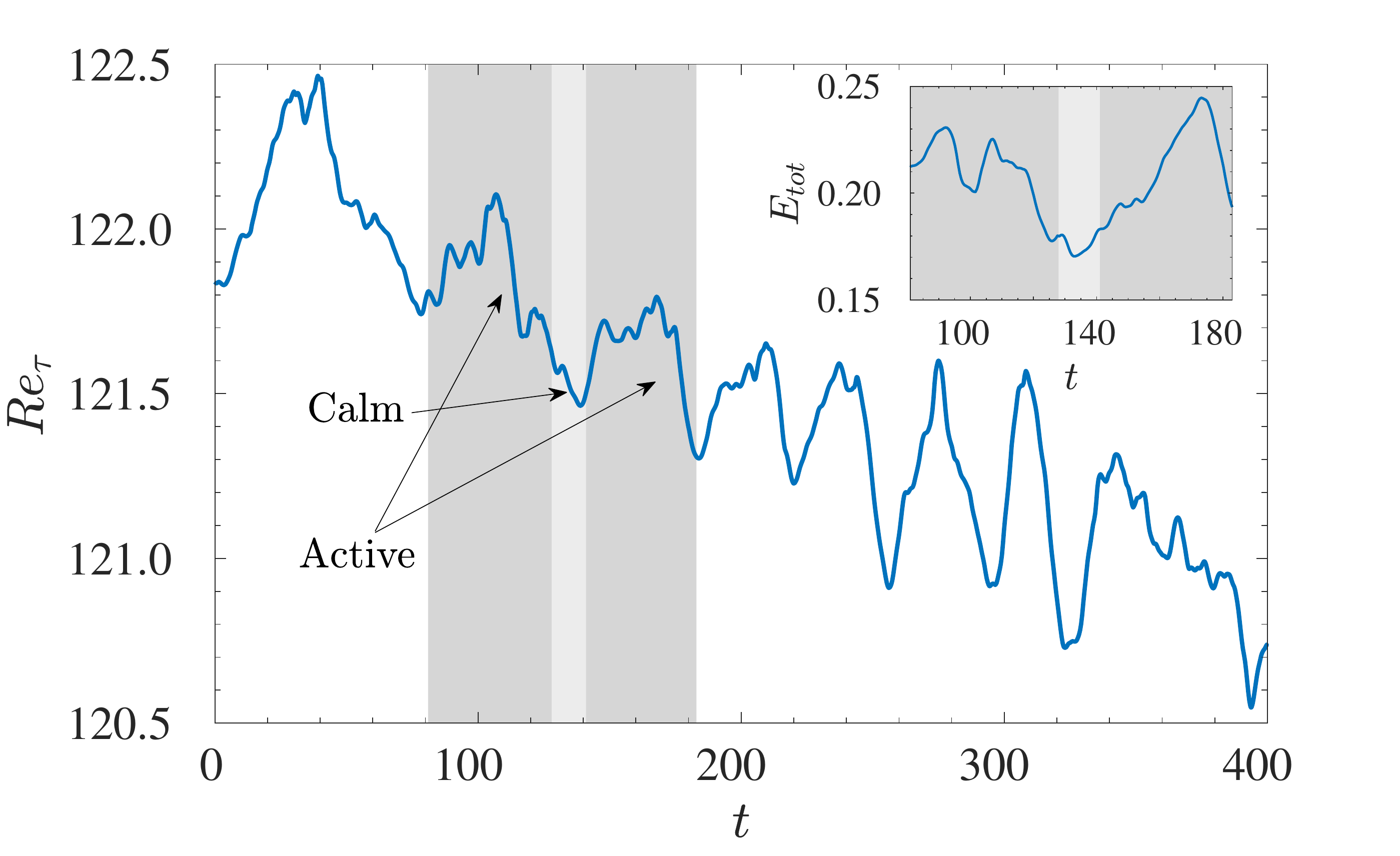}
	\caption{Time history of the friction Reynolds number ($\Rey_\tau$) calculated over the whole computational domain, and of the perturbation kinetic energy ($E_{tot}$) calculated on a sub-domain co-moving with the upstream front of the slug at $Re=3300$. Dark and light shaded areas highlight examples of active and calm phases at the front.}
	\label{fig:timemodul}
\end{figure}

If the pipe is bent, the upstream front gradually emerges from the laminar flow and is considerably less densely populated by vortical structures.
The generation of new vortices is modulated in time and is organised over the succession of ``active'' and ``calm'' phases, see figures~\ref{fig:lambda2}~{(b)--(c)} and supplementary videos 3--5.
This is particularly evident for moderately expanding slugs.
During the active phase new vortices are generated at the outer bend, where $P_k$ is maximum and perturbations are amplified, as discussed in \S\ref{sec:tke_bud}.
During the calm phase vortical structures are generated at a lower rate near the centre of the pipe.
The alternation of calm and active phases is illustrated in figure~\ref{fig:timemodul}, which reports the time history of the friction Reynolds number and of the total perturbation kinetic energy.
The former is defined as $\Rey_\tau = \left( \Rey / \nu \right) \sqrt{ \left( R / 2\rho \right) \left( \mathrm{d}p / \mathrm{d} s \right) \, }$, and is computed using the mean pressure gradient calculated over the entire domain.
The latter reads $E_{tot} = \left(1/2\right) \int_{Vol} \sum \Delta u_i^2 \mathrm{d}V$, with $\Delta u_i = u_i - U_i$, and is calculated on a 20-diameter long moving frame of reference centred on the upstream front of the slug.
The signal of $\Rey_\tau$ shows a decreasing trend that is consistent with the fact that a fully turbulent flow at $\Rey=3300$ is characterised by sub-laminar drag~\citep[see][]{Cioncolini2006,Noorani2015}.
Active and calm phases appear as local oscillations on top of this trend.
In particular, active phases correspond to a local plateau and subsequent decrease in the profile of $\Rey_\tau$, and are connected by calm phases where $\Rey_\tau$ reaches a local minimum before starting to rise again.
Active phases typically last for a time $t = \mathcal{O}(40 \, D/U)$, while calm phases are shorter, $t = \mathcal{O}(20 \, D/U)$.
Since $\Rey_\tau$ is based on an integral quantity calculated over the whole computational domain, which is filled by the turbulent slug as time advances, the effect of the dynamics at the front is gradually less visible.
The perturbation kinetic energy calculated across the slug front shows that active phases consistently correspond to high values of $E_{tot}$.

In addition to the spatial localisation and lower values of $P_k$ discussed in \S\ref{sec:tke_bud}, the temporal modulation also contributes to weaker upstream fronts if compared to straight pipes.
To the best of our knowledge, this is the first time this phenomenon is documented on the fronts of expanding turbulence in pipe flows.

The curvature also influences the transport of vortices inside the slug.
This can be clearly seen in the supplementary video 1, which presents a comparison between a straight and a bent pipe.
The video shows vortical structures in a one-diameter long section as seen from inside the pipe while following the upstream front.
While in the straight pipe the vortices move chaotically, in the bent pipe they are clearly transported in a coherent circulatory motion dictated by the Dean vortices.
This visually shows how a localised production of vortices is sufficient to sustain the extended structure of a slug.

%% file: conclusions.tex
We have studied the dynamics of localised turbulent structures in a bent pipe with curvature $0.01$.
The analysis was performed by means of direct numerical simulations for Reynolds numbers between $2900$ and $5000$.
Our results show that subcritical transition to turbulence occurs over $2950 \lesssim \Rey \lesssim 3100$, where laminar and turbulent flow coexist in an intermittent fashion.
The indicated range is significantly narrower than the one documented for a straight pipe, which is in good agreement with the literature on the subject \citep[see, \eg][]{Sreenivasan1983,Kuhnen2015}.

Localised turbulent structures in bent pipes bear qualitative similarities to those in straight pipes in that they appear in the form of puffs and slugs that are sustained by an instability at their upstream front.
However, the most striking difference is that the strong upstream front vanishes if the pipe is bent.
We report velocity fluctuations that are three times smaller than the ones observed in straight pipes.
Furthermore, the turbulent kinetic energy budget shows that production at the front is at most one-and-a-half times the one in the core of the slug, while in straight pipes it is up to five times larger than in the core.
Production and dissipation are strongly localised towards the outer bend, in a region that is highly receptive to flow perturbations.
We argue that the emergence of a strong secondary flow~\citep{Dean1927} is responsible for the localisation and is crucial for the effective mixing of vortical structures.
A further peculiarity encountered in bent pipes is that the generation of vortices at the front occurs through a succession of active (longer) and calm (shorter) phases, which contribute to the overall small values of production and dissipation.
This is to the best of our knowledge the first time that such temporal modulation of the upstream front is reported in pipe flows.

The final difference between bent and straight pipes is the apparent absence of puff splitting when the pipe is bent, at least for the choice of parameters investigated here.
Since the genesis of a child puff from a mother puff is a statistical (non-deterministic) phenomenon, we cannot completely exclude it based on our limited number of simulations.
A thorough, possibly experimental, investigation focused on this phenomenon should be performed to verify our claim.
We conjecture that the absence of puff splitting is connected to the weakness and cross-sectional localisation of the upstream fronts.
Turbulent structures protruding from the downstream front of the mother puff into the surrounding laminar flow~\citep{Avila2011} have a low probability of entering the small region of high amplification located near the outer wall.
Moreover, due to the secondary motion, the few vortical structures that visit this region do not linger for long enough to sustain a new puff after the separation between mother and daughter~\citep{Shimizu:2014bu}.

\bigskip
Financial support from the Knut and Alice Wallenberg Foundation via the Wallenberg Academy Fellow programme is gratefully acknowledged.
Simulations and post-processing were performed on resources provided by the Swedish National Infrastructure for Computing (SNIC)  at PDC (Beskow and Tegner).
The authors would like to acknowledge M.\ Schanen and O.\ Marin (Argonne National Laboratory) for their support in developing the nonlinear optimisation routines.